\documentclass[hyper]{JHEP3}

\usepackage{cite}

\usepackage{epsfig}
\usepackage{amsmath,mathenv}

\usepackage{latexsym,exscale}

\newcommand\fverb{\setbox\fverbbox=\hbox\bgroup\verb}
\newcommand\fverbdo{\egroup\medskip\noindent%
			\fbox{\unhbox\fverbbox}\ }
\newcommand\fverbit{\egroup\item[\fbox{\unhbox\fverbbox}]}
\newbox\fverbbox

\newcommand{\rig}{\rightarrow}

\newcommand{\be}{\begin{eqnarray*}}
\newcommand{\ee}{\end{eqnarray*}}
\newcommand{\gl}[1]{(\ref{#1})}

\newcommand{\bee}{\begin{eqnarray}}
\newcommand{\eee}{\end{eqnarray}}
\newcommand{\beeq}{\begin{equation}}
\newcommand{\eeeq}{\end{equation}}

\def\slashed#1{#1\llap{\sl/}}

\def\citere#1{\mbox{Ref.~\cite{#1}}}
\def\citeres#1{\mbox{Refs.~\cite{#1}}}

\def\MW{M_{\mathrm{W}}}
\def\MZ{M_{\mathrm{Z}}}
\def\pT{p_{\mathrm{T}}}
\def\pTx#1{p_{\mathrm{T,#1}}}
\newcommand{\dsl}[1]{\not \hspace{-0.7mm}#1}

\title{NLO QCD corrections to WZ+jet production with leptonic decays}

\author{
Francisco Campanario\\
Institute for Theoretical Physics, 
Karlsruhe Institute of Technology,\\
76128 Karlsruhe, Germany}
\author{
Christoph Englert\\
Institute for Theoretical Physics, 
Karlsruhe Institute of Technology,\\
76128 Karlsruhe, Germany \\
and\\
Institute for Theoretical Physics,
Heidelberg University,\\
69120 Heidelberg, Germany}
\author{
Stefan Kallweit\\
Paul Scherrer Institute, W\"urenlingen and Villigen\\
5232 Villigen PSI, Switzerland}
\author{
Michael Spannowsky\\
Institute of Theoretical Science, University of Oregon,\\
Eugene, OR 97403-5203, USA}
\author{
Dieter Zeppenfeld\\
Institute for Theoretical Physics, 
Karlsruhe Institute of Technology,\\
76128 Karlsruhe, Germany}

\abstract{
We compute the next-to-leading order QCD corrections to WZ+jet production at the Tevatron and the 
LHC, including decays of the electroweak bosons to light leptons with all off-shell effects taken into account. 
The corrections are sizable and have significant impact on the differential distributions.
}


\preprint{KA-TP-11-2010\\PSI-PR-10-09\\SFB/CPP-10-41}

\begin{document}
%
%
%
\section{Introduction}
At the Large Hadron Collider (LHC) as well as at the Tevatron, electroweak di-boson production in association with a hard jet
represents an important class of processes, of either signal or background character in various searches
for Standard Model~(SM) and beyond. The rates are large, especially at the LHC by accessing the gluon
density at small momentum fraction, and next-to-leading order~(NLO) QCD corrections have
turned out sizable in a series of recent publications \cite{Dittmaier:2007th,Dittmaier:2009un,Campbell:2007ev,Campanario:2009um,
Binoth:2009wk} providing WW+jet, ZZ+jet, and ${\rm{W^\pm\gamma}}$+jet production at NLO QCD precision at hadron colliders.
In this paper, we supplement NLO QCD precision to 
\be
p\bar{p},pp \rig 3\,\rm{leptons}~+\dsl{E}_{\mathrm{T}}+ {\rm jet} + \mathrm{X}\,,
\ee
i.e. to W$^\pm$Z+jet production including full leptonic decays. 
We give cross sections for LHC and Tevatron collisions and also discuss the corrections' phase-space dependence
by investigating differential correction factors at the LHC.

To verify our results, special care is devoted to independent numerically stable implementations of the processes, yielding two independent
fully-flexible Monte Carlo programs, based on different approaches.

We organize this work in the following way: In section \ref{sec:details} we provide an overview of the two programs
we have employed for the numerical results of this paper, to which section \ref{sec:numerics} is devoted. We first focus
on on-shell production at the Tevatron and the LHC in sections \ref{sec:tevaresults} and \ref{sec:lhcresults}, respectively, and
then move on to discuss the differential impact of the QCD corrections for the LHC setup in more phenomenological
detail in section \ref{sec:distriblhc}. Section \ref{sec:summary} concludes with a summary of the work presented in this
paper.

\section{Details of the calculation}
\label{sec:details}
We invoke a dedicated system of checks and balances to validate our calculation. In particular, we have compared two different Monte Carlo
implementations, relying on distinct approaches. The comparisons involve cross checks at the amplitude level for a fixed phase-space point
as well as comparisons of integrated cross sections. In both cases we find agreement on the level of double-precision accuracy or agreement within the statistical errors, respectively,
for different choices of renormalization and factorization scales and cuts.

Our independent calculations are based on the approaches of \cite{Dittmaier:2007th,Dittmaier:2009un,Campanario:2009um}, where
the used methods have already been described in detail. Hence, we limit ourselves to the bare necessities to make this paper self-consistent
and refer the interested reader to the above publications for more details.\\[0.2cm]
\begin{figure}[!t]
\begin{center}
\includegraphics[scale=0.4]{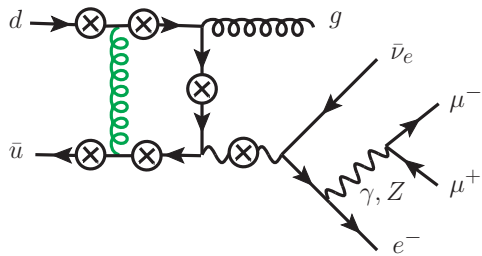}
\caption{\label{feyngraph} Representative Feynman graph contributing to $p\bar p, pp \rig \bar\nu_e e^-\mu^+\mu^-+\mathrm{X}$. The crosses 
mark points where the $\gamma,Z\rig\mu^+\mu^-$ decay topology can be inserted. Indicated is also the exchange of a virtual gluon, which gives rise to
self-energy, triangle, box, and pentagon topologies. Not shown are topologies that result from closed fermion loops, non-abelian graphs, and real emission topologies.
All other subprocesses can be recovered from the shown graph by flavour summation and/or crossing.}
\end{center}
\end{figure}
\underline{\emph{Program 1}}\quad We generalize the NLO QCD calculation of ${\rm{W\gamma}}$+${\rm{jet}}$ production (including leptonic decays) \cite{Campanario:2009um} to
WZ+${\rm{jet}}$ production. The leading order~(LO) matrix elements at ${\cal{O}}(\alpha^4\alpha_{\rm{s}})$, cf. figure \ref{feyngraph}
are calculated with {\sc Helas} routines  \cite{Murayama:1992gi} 
generated with {\sc MadGraph} \cite{Alwall:2007st}. Our phase-space implementation is based on routines readily present in the 
{\sc Vbfnlo} suite \cite{Arnold:2008rz}, which were already applied in the context of NLO QCD vector-boson-fusion WZ+$2{\rm{jets}}$ production, 
including leptonic decays, in various scenarios \cite{Bozzi:2007ur,Englert:2008wp}. Nonetheless integrated results for the different subprocesses 
have been checked against {\sc Sherpa}~\cite{Gleisberg:2008ta}, yielding agreement within statistical errors on per-mill level.

The  virtual corrections are combined to groups that include all loop diagrams derived from
a born-level configuration, i.e. all self-energy, triangle, box and pentagon corrections to a quark line with three attached gauge bosons, which
are computed as effective decay currents in case of the electroweak bosons, 
are combined to a single numerical routine. This leaves a set of universal building blocks, which were already appropriately assembled
to determine the one-loop contribution to ${\rm{W\gamma}}$+${\rm{jet}}$ production (cf. \cite{Campanario:2009um} for details on the verification
of the implementation against an independent approach). The generalization to WZ+${\rm{jet}}$ becomes trivial by replacing the photon
polarization vector by the effective Z decay current multiplying the appropriate coupling. These building blocks are set up using 
in-house routines within the framework of {\sc FeynCalc} \cite{Mertig:1990an} and {\sc FeynArts} \cite{Hahn:2000kx}. They invoke the
Passarino--Veltman reduction \cite{Passarino:1978jh} up to boxes and the Denner--Dittmaier reduction \cite{Denner:2002ii} for pentagons.
The remaining fermionic loop corrections
are derived via algebraic calculations using {\sc FeynCalc}, which is subsequently processed to {\sc Fortran} routines with in-house routines. The 
scalar integrals that are not already present in the {\sc Vbfnlo} framework are supplemented from the Ellis--Zanderighi library \cite{Ellis:2007qk}.
All effective decay currents are evaluated by means of {\sc Helas} routines generated with {\sc MadGraph}, which are modified to fit our purpose of calculating the one-loop amplitude.

To speed up the numerical implementation of the numerous subprocesses that show up as part of the real emission, we computed the real emission
matrix element using the spinor helicity formalism of \cite{Hagiwara:1988pp}. We store intermediate numerical results common to all subprocesses 
and re-use them whenever possible. All matrix elements have been checked explicitly against code generated from {\sc MadGraph}.

The infrared~(IR) singularities are subtracted applying the dipole subtraction of \cite{Catani:1996vz}, while the corresponding
LO matrix elements and currents for the subtraction kinematics are computed using {\sc Helas} routines. We also apply necessary
bookkeeping in order not to waste computing time.
The IR poles of the virtual amplitude are cancelled against the real-emission ones algebraically,
and we perform the integration of finite collinear terms as part of the real-emission integration by appropriately mapping the born-type configuration
as done in \cite{Figy:2007kv}.

The code will become publicly available with an upcoming update of {\sc Vbfnlo}.\\[0.2cm]
\underline{\emph{Program 2}}\quad
We proceed essentially in the same way as in the calculation of WW+jet, 
which is discussed in some detail in \citeres{Dittmaier:2007th,Dittmaier:2009un}.
All LO helicity amplitudes are calculated by application of the 
Weyl--van-der-Waerden formalism (as described in \citere{Dittmaier:1998nn}). 
In this approach, the implementation of the gauge-boson decays can be easily 
realized by replacing the polarization bispinors of the gauge bosons with 
expressions containing the currents of the decay leptons. 
Allowing for off-shell gauge bosons and still respecting gauge invariance 
requires the inclusion of 
diagrams which do not contain two simultaneously resonant gauge-boson propagators. 
However, all diagrams of this kind can be constructed from W+jet-production 
amplitudes by replacing the W-boson polarization bispinor with an appropriate expression 
describing its decay into 4 leptons. Additionally, diagrams with intermediate 
photons instead of Z bosons have to be taken into account. It is worth noting 
that the described replacements are exactly the same for the 
LO and for all components of the NLO QCD calculation---in other words, 
the bispinor replacements are universal. In particular, 
no new types of loop diagrams show up in the virtual corrections.

Again, the dipole subtraction formalism of \cite{Catani:1996vz} is applied 
to rearrange the IR divergences between real and virtual corrections at NLO QCD.

The loop diagrams and amplitudes are generated by 
{\sc FeynArts}~3.4 \cite{Hahn:2000kx} and then further manipulated with {\sc FormCalc}~6.0
\cite{Hahn:1998yk} to automatically produce {\sc Fortran} code.
The whole reduction of tensor to scalar integrals is done with the
help of the {\sc LoopTools} library \cite{Hahn:1998yk},
which also employs the Denner--Dittmaier method~\cite{Denner:2002ii} for the
5-point tensor integrals, Passarino--Veltman \cite{Passarino:1978jh}
reduction for the lower-point tensors, and the {\sc FF} package 
\cite{vanOldenborgh:1989wn,vanOldenborgh:1991yc} for the evaluation 
of regular scalar integrals. The IR~(soft and collinear) singular 3- and 4-point
 integrals in dimensional regularization are linked to this library 
as in the WW+jet calculation. Again, the explicit results of 
\citere{Dittmaier:2003bc} for the vertex and of \citere{Bern:1993kr} 
for the box integrals (with appropriate analytical continuations) are taken. 
Actually the {\sc FormCalc} package assumes a four-dimensional
regularization scheme for IR divergences, i.e.\ rational terms of IR
origin are neglected by {\sc FormCalc}. However, in \citere{Bredenstein:2008zb} 
it was generally shown that such rational terms consistently cancel if UV and 
IR divergences are properly separated. 
Thus we could use the algebraic result of {\sc FormCalc} for the unrenormalized 
amplitudes without any modification, apart from supplementing the needed
IR-singular scalar integrals.

To receive the real-correction matrix elements we also employ the 
Weyl--van-der-Waerden formalism. The dipoles needed to cancel the 
divergences in the respective subprocesses are automatically generated from 
the born-level helicity amplitudes.
To achieve numerical stability on a high-accuracy level, the phase-space 
integration is performed by a multi-channel Monte Carlo 
integrator~\cite{Berends:1994pv} with weight optimization~\cite{Kleiss:1994qy}, 
which has been written in {\sc C++} and checked in detail in the calculation 
of WW+jet.
Additional channels basing on dipole kinematics are automatically 
included to improve the efficiency of the integration of the 
dipole-subtracted real-emission matrix elements. 
When the full calculation with off-shell gauge bosons is considered, only 
channels according to doubly-resonant diagrams are included, which turns out to 
provide already sufficient numerical stability.\\[0.2cm]

Both numerical programs account for finite width effects of the electroweak gauge bosons 
(when considering their leptonic decays) with a fixed-width scheme, which is also the scheme 
used by {\sc MadGraph}: while we calculate with Breit-Wigner propagators of the W and Z bosons  
we keep the weak mixing angle real. 
To justify this approach, which breaks gauge invariance, we compared the results to a calculation 
performed applying the complex-mass scheme~\cite{Denner:1999gp} 
in one of the programs. We find an agreement on the per-mill level between the two calculations, 
so the effect of gauge-invariance breaking turns out to be sufficiently small to be ignored here.

\section{Numerical results}
\label{sec:numerics}
Throughout, we use CTEQ6M parton distributions \cite{Pumplin:2002vw}
at NLO, and the CTEQ6L1 set at LO. We choose $\MZ=91.1876~\rm{GeV}$, $
\MW=80.425~\rm{GeV}$, and $G_F=1.16637\times 10^{-5}~\textnormal{GeV}^{-2}$ as electroweak input 
parameters and derive the electromagnetic coupling $\alpha$ and the weak mixing 
angle $\sin\theta_{\rm{w}}$ via SM tree-level relations. The LO and NLO
running of $\alpha_{\rm{s}}$ are determined by 
$\alpha_s^{\rm{LO}}(\MZ)=0.130$ and $\alpha_s^{\rm{NLO}}(\MZ)=0.118$ for five active flavors, respectively.

The center-of-mass energy is fixed to $\sqrt{s}=14~\rm{TeV}$ for LHC and $\sqrt{s}=1.96~\rm{TeV}$ for Tevatron collisions, respectively. 
We consider both on-shell production of the electroweak bosons and their decays to distinct species of light leptons, 
e.~g.~\mbox{$W^-\rightarrow e^-\bar \nu_e$} and 
\mbox{$Z\rightarrow \mu^-\mu^+$}, treating these leptons as massless. The CKM matrix is taken to be diagonal, and we 
neglect bottom contributions throughout because they are numerically 
negligible anyway and can even be further suppressed by b-tagging. To be more precise, we neglect the---finite and negligibly small---contribution 
from real correction minus subtraction 
terms if external bottom quarks are involved. In the fermionic quark loops and, correspondingly, the $I$-operator, 
we keep all six quark flavours.
A non-diagonal CKM matrix decreases our LHC results only at the per-mill level because gluon-induced subprocesses 
dominate the cross section. In case of a Cabibbo-like block-diagonal CKM matrix,  the contribution from this subset of subprocesses 
is not affected if all light quarks are summed over.
The correction for the Tevatron cross section is about 3\% due to the dominance of quark-induced subprocesses. These 
corrections are well below 
the residual scale dependence at NLO QCD.
The final-state partons are recombined to massless jets via the algorithm of \cite{Ellis:1993tq} with resolution 
parameter $R=1.0$. Other jet algorithms, like 
the kT algorithm of Ref.~\cite{Catani:1993hr} have also been implemented in program 1.

\subsection{Event selection}
\label{sec:eventsel}
To analyze the impact of the NLO QCD corrections on the total production cross sections at both the Tevatron and the LHC, we apply
a rather inclusive set of cuts. 
In case of on-shell W and Z boson production, the jets are required to have a transverse momentum of
\bee
\label{eq:ptcut}
\pTx{jet}\geq 50~\rm{GeV}\,,
\eee 
which is the only selection criterion we impose for the calculation of the cross sections of Secs. \ref{sec:tevaresults} and \ref{sec:lhcresults}.
In case of included leptonic decays of the electroweak bosons, we account for finite jet-detection coverage by requiring the jets
to have rapidities
\bee
|\eta_j|\leq 4.5
\eee
in addition to the cut on $\pTx{jet}$ of Eq.~\gl{eq:ptcut}. However, with the used cut value for $\pTx{jet}$, the additional effect of this 
cut is completely negligible at the given collider energies.
All leptons are required to lie in 
\bee
|\eta_\ell|\leq 2.5
\eee
with transverse momenta of
\bee 
\pTx{\ell} \geq 25~\rm{GeV}\,.
\eee
The overall missing transverse momentum is chosen to be 
\bee
\slashed{p}_{\rm{T}} \geq 25~\rm{GeV}\,.
\eee
The leptons have to be separated in the azimuthal-angle--pseudorapidity plane by 
\bee
R_{\ell\ell'} = (\Delta \phi_{\ell\ell'}^2 + \Delta\eta_{\ell\ell'}^2)^{1/2}\geq 0.2\,,
\eee 
and for their separation from
observable jets we choose 
\bee
R_{\ell j}\geq 0.4\,.
\eee
It is customary from a theoretical point of view to study also the cross sections' behaviour with a veto applied to the second resolved jet in the context
of mono-jet production: it was shown \cite{Campbell:2007ev,Dittmaier:2007th,Dittmaier:2009un,Campanario:2009um,Binoth:2009wk} that this additional veto
yields highly stabilized NLO exclusive cross sections in the context of WW+jet, ZZ+jet, and W$\gamma$+jet production. Indeed, given the similarities of the processes from
a QCD point of view, identical properties for WZ+jet production are evident. 
We will discuss the phenomenological problems of the additional jet veto in detail in section \ref{sec:distriblhc}. 
\begin{figure}[!t]
\centering
\includegraphics[bb = 170 410 450 660, scale = .8]{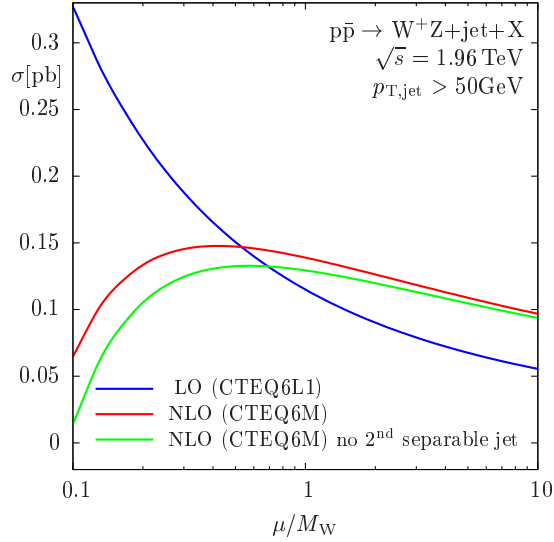}
\caption{\label{fig:scvarmwteva} Fixed-scale variation of $\mu_{\rm{R}}=\mu_{\rm{F}}=\mu$ for on-shell W$^\pm$Z+jet production at the Tevatron.}  
\end{figure}

\subsection{Production cross sections at the Tevatron}
\label{sec:tevaresults}
For on-shell W$^\pm$Z+jet production from Tevatron collisions with the additional requirement of \gl{eq:ptcut}, we compute a total inclusive and exclusive
WZ+jet on-shell cross sections and $K$ factors
\bee
K={\sigma^{\rm{NLO}}\over\sigma^{\rm{LO}}}
\eee
of 
\bee
\sigma^{\rm{NLO}}_{\rm{incl}}({\rm{W^\pm Z+jet}})=(139.01\pm 0.10)~{\rm{fb}} \quad \left[ K=1.209 \right]\,, \\
\sigma^{\rm{NLO}}_{\rm{excl}}({\rm{W^\pm Z+jet}})=(129.40\pm 0.10)~{\rm{fb}} \quad \left[ K=1.125\right]\,,
\eee
which are
dominated by $q\bar Q$ induced processes due to the relatively large momentum fraction of the incoming partons 
$x\sim 0.2$ at LO, which we infer from the Monte Carlo simulation.
The total correction of $21\%$
with respect to LO at the central scale is sizable.
Nonetheless, including the leptonic decays decreases the cross sections
to phenomenologically subdominant size, unless the transverse momentum requirement for jets is reduced substantially. 
We therefore limit ourselves to quoting total on-shell production
rates at the Tevatron and focus on differential distributions at the LHC only. 

A lower bound on the scale uncertainties of the cross sections can be inferred e.~g. from varying the fixed renormalization and factorization scales by a factor
of two around the central value $\mu_{\rm{R}}=\mu_{\rm{F}}=\MW$, cf. figure \ref{fig:scvarmwteva}. Doing so, the LO approximation exhibits a scale variation of 
$31\%$ which is decreased to $9\%$ by including NLO-inclusive precision.

\subsection{Production cross sections at the LHC}
\label{sec:lhcresults}
Turning to the more energetic LHC collisions, we find a completely different situation
compared to the Tevatron. The proton is typically probed at much lower momentum fractions $x\sim 0.02$ at LO (as the Monte Carlo calculation shows), so that the $qg$-induced initial states dominate the total rate. The total NLO-inclusive cross sections are
\bee [llqql]
\sigma^{\rm{NLO}}_{\rm{incl}}({\rm{W^-Z+jet}}) &= (7.495\pm 0.008)~{\rm{pb}} &  \left[ K({\rm{W^-Z+jet}})=1.298 \right] \,, \\
\sigma^{\rm{NLO}}_{\rm{incl}}({\rm{W^+Z+jet}})&= (12.061\pm 0.013)~{\rm{pb}} & \left[ K({\rm{W^+Z+jet}})=1.260 \right] \,,
\eee
\begin{figure}[!t]
\centering
\includegraphics[bb = 170 410 450 660, scale = .8]{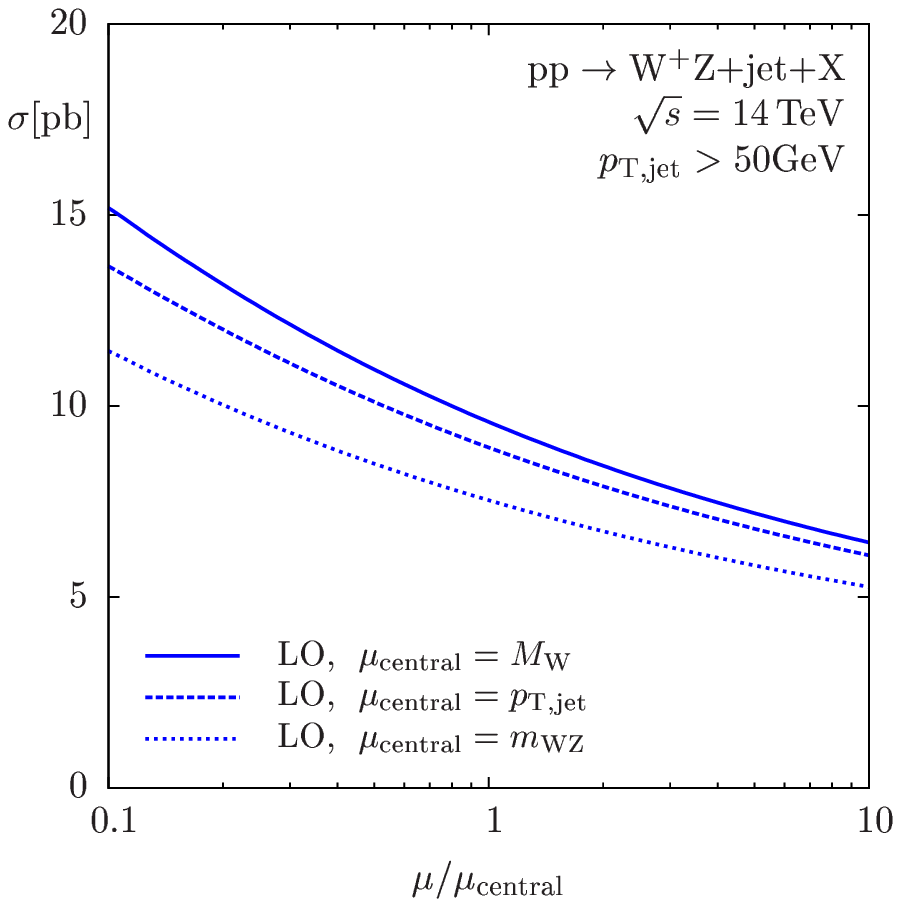}
\includegraphics[bb = 160 410 410 660, scale = .8]{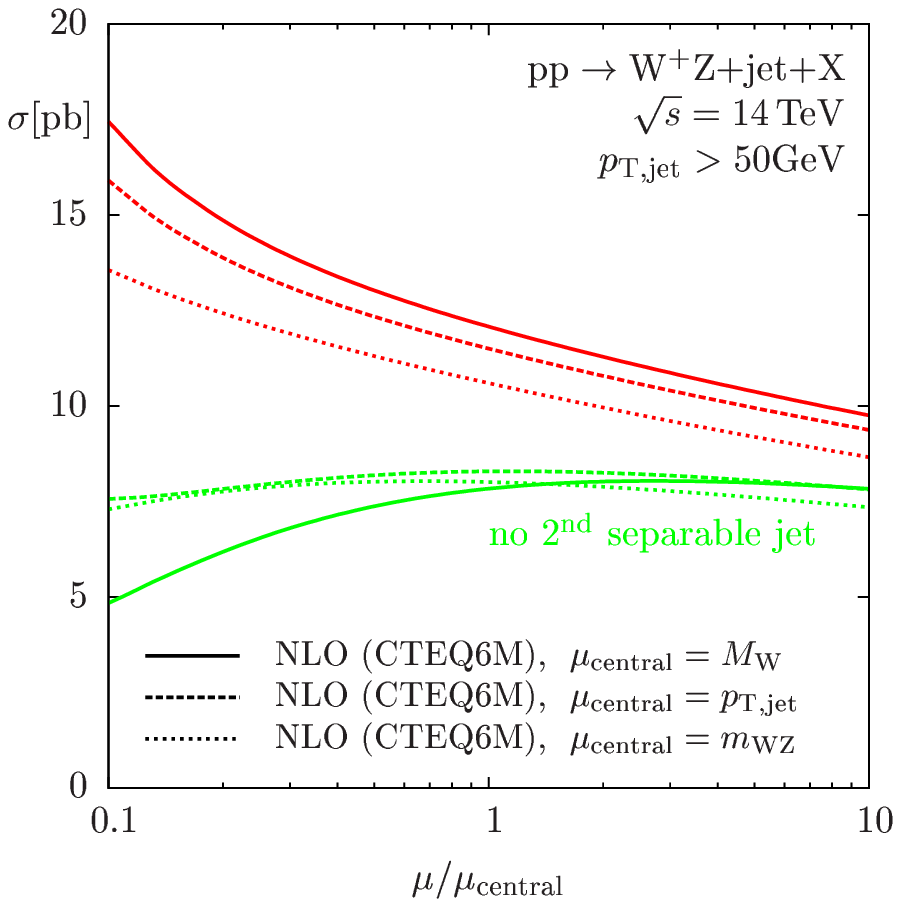}\\[2ex]
\includegraphics[bb = 170 410 450 660, scale = .8]{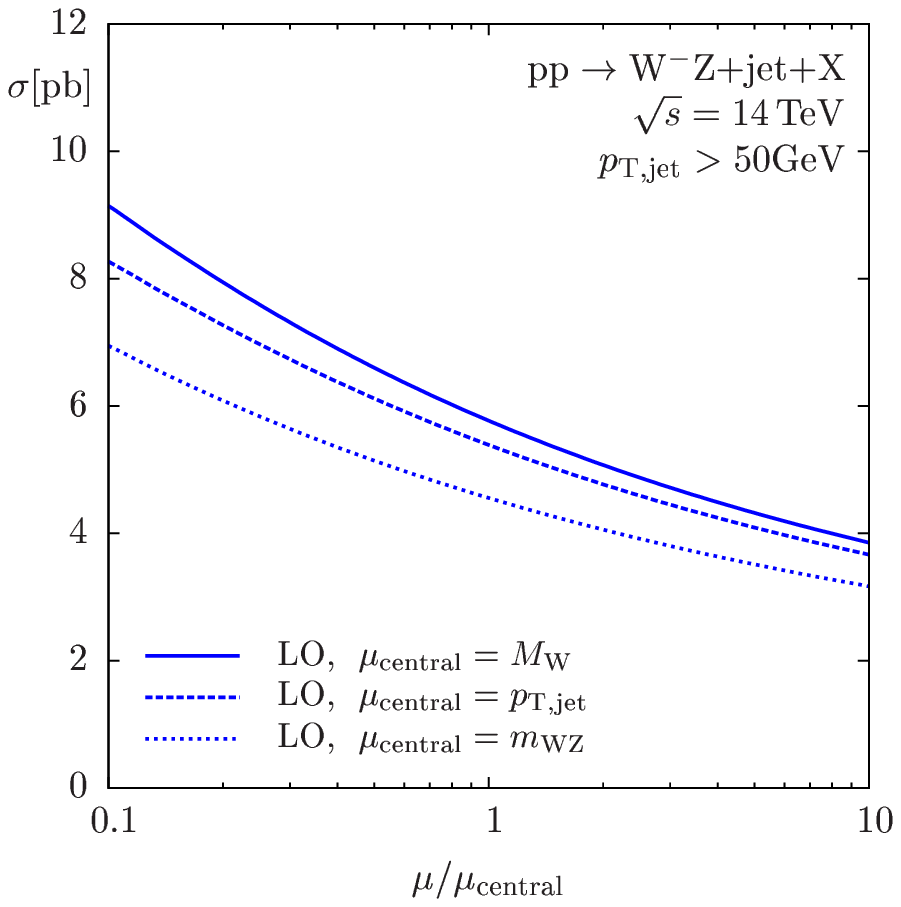}
\includegraphics[bb = 160 410 410 660, scale = .8]{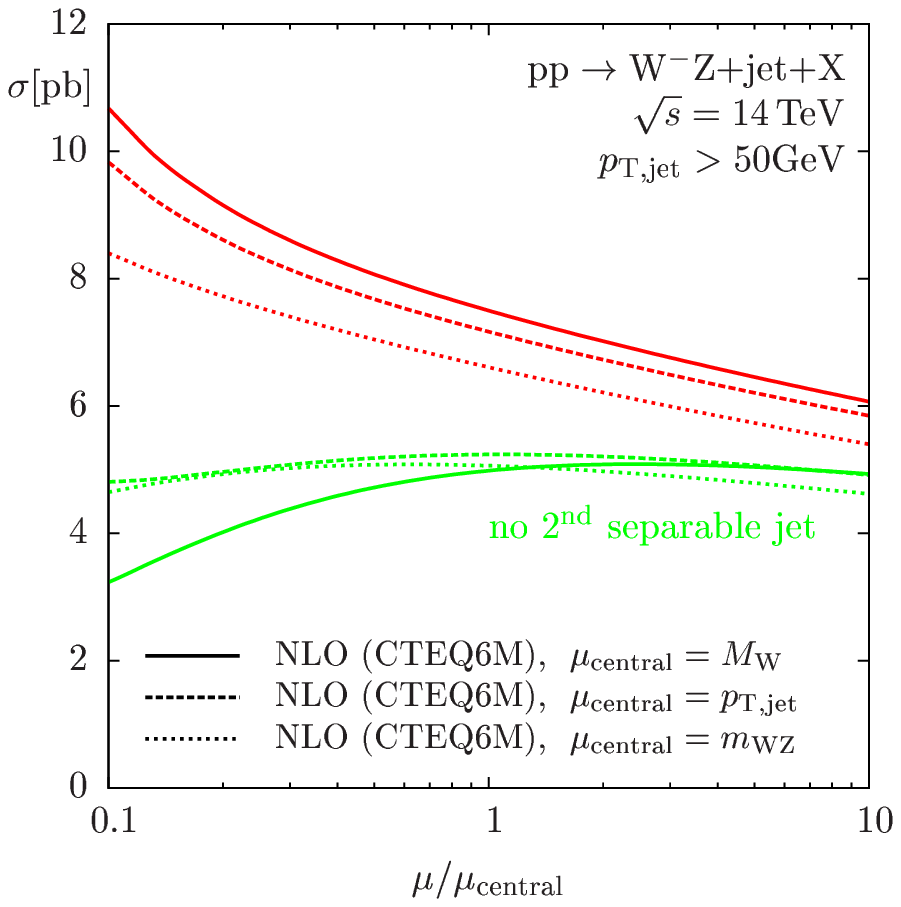}\\
\caption{\label{fig:scvarlhc} Scale variation for on-shell W$^+$Z+jet (upper plots) and W$^-$Z+jet (lower plots) production at the LHC. The identified renormalization and factorization scales $\mu_{\rm{R}}=\mu_{\rm{F}}=\mu$ are varied with respect to the fixed scale $\MW$, the maximum $\pTx{jet}$, and the invariant mass $m_{\rm{WZ}}$. The plots on the left show cross sections at LO, those on the right at NLO QCD accuracy.}  
\end{figure}
for a scale choice $\mu_{\rm{R}}=\mu_{\rm{F}}=\MW$. 

In close analogy to \cite{Campbell:2007ev,Dittmaier:2007th,Dittmaier:2009un,Campanario:2009um,Binoth:2009wk}, the di-jet
contribution re-introduces a substantial dependence on the renormalization scale $\mu_{\rm{R}}$ via the dominating $q g$-induced channels. 
This becomes apparent by 
checking the variation for several scales intrinsic to the total cross section in figure \ref{fig:scvarlhc}.

The cross sections' qualitative scaling behaviour does not depend on the choice of the intrinsic scale, 
and the characteristic 
increase of the NLO-inclusive cross sections at small scales $\mu=\mu_{\rm{R}}=\mu_{\rm{F}}$ reflects the renormalization
scale dependence of the di-jet contribution, which is a leading order-$\alpha_{\rm{s}}$ contribution to our NLO computation.

This $\mu_{\rm{R}}$ dependence of di-jet contributions can be effectively buffered by imposing an additional veto on events with two resolved jets, which
gives rise to total NLO-exclusive rates for $\mu_{\rm{R}}=\mu_{\rm{F}}=\MW$ of
\bee[llqql]
\label{eq:vetoonshell}
\sigma^{\rm{NLO}}_{\rm{excl}}({\rm{W^-Z+jet}}) &= (4.981\pm 0.009)~{\rm{pb}} &  \left[ K({\rm{W^-Z+jet}})=0.862\right] \,,\\
\sigma^{\rm{NLO}}_{\rm{excl}}({\rm{W^+Z+jet}})&= (7.831\pm 0.014)~{\rm{pb}} & \left[ K({\rm{W^+Z+jet}})=0.818\right] \,.
\eee
Varying again $\mu_{\rm{R}}=\mu_{\rm{F}}$
by a factor two around the central values in figure \ref{fig:scvarlhc}
amounts to scale uncertainties of $8\%$ (W$^-$Z+jet) and $8\%$ (W$^+$Z+jet) of the total inclusive cross sections at the LHC. For the vetoed sample, the scale dependence is 
reduced to about $5\%$ for W$^-$Z+jet, and $6\%$ for W$^+$Z+jet production.

\begin{figure}
\centering
\includegraphics[bb = 200 380 450 700, scale = .8]{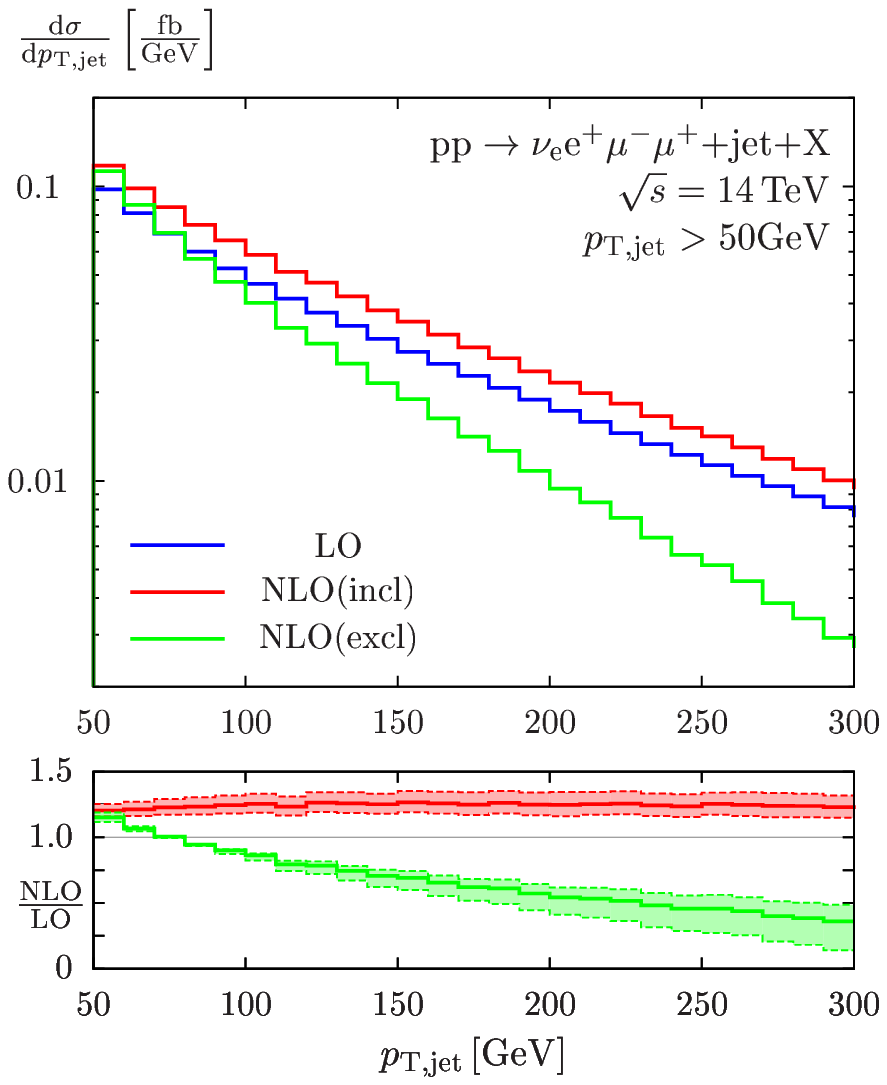}
\includegraphics[bb = 160 380 410 700, scale = .8]{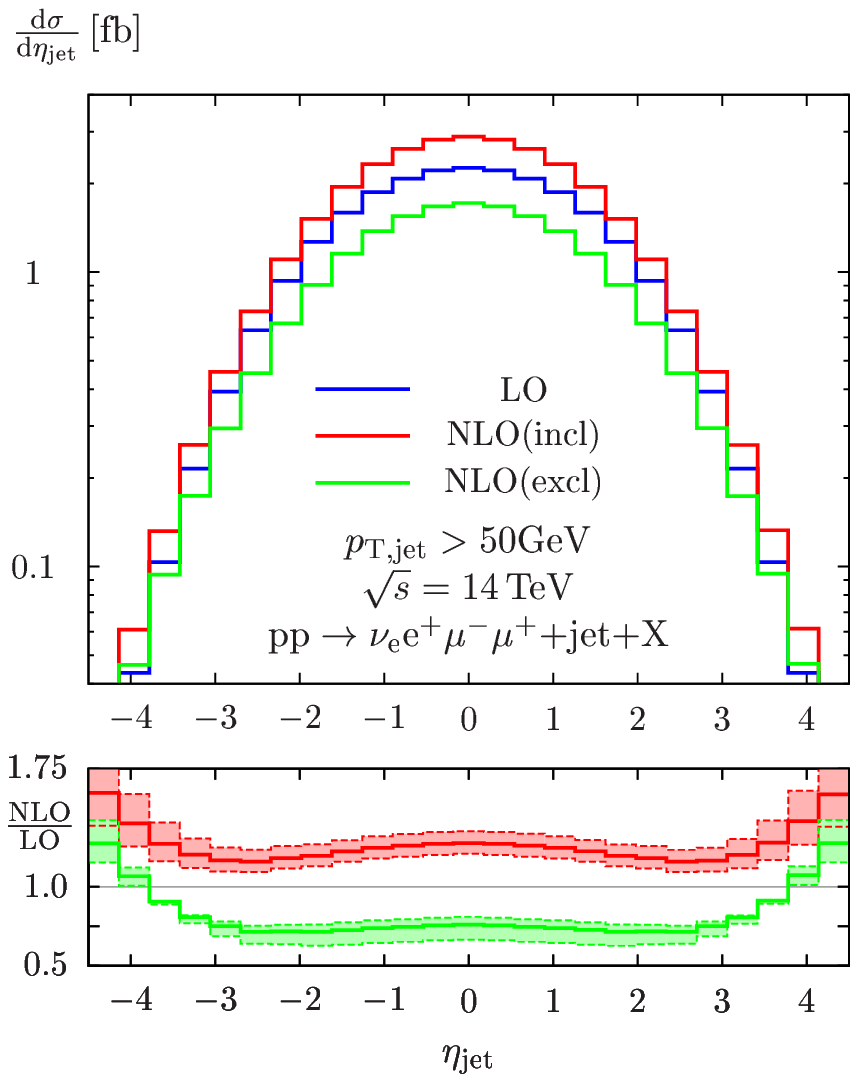}\\[-2em]
\caption{\label{fig:ptjetaj} LO, NLO-inclusive and NLO-exclusive differential distributions of the transverse momentum of the hardest 
jet and its pseudorapidity for W$^+$Z+jet production including leptonic decays. 
The differential $K$-factor band corresponds to varying $\mu_{\rm{R}}=\mu_{\rm{F}}$ by a factor two around the central scale
in the NLO distribution only.}
\end{figure}

The improved perturbative stability of the exclusive cross sections should be interpreted with caution. 
While jet-vetoing is a straightforward exercise in the context of fixed-order Monte Carlo calculations, 
its phenomenological consequences are generally highly delicate, both from the theoretical and the experimental side.
The small total correction along with the stability against variations of $\mu_{\rm{R}}=\mu_{\rm{F}}$ of the exclusive cross
sections should therefore not be misinterpreted as a guideline to stable LHC predictions {\it per se}, but as a significant perturbative
improvement of WZ+jet production up to the specified threshold value of $\pTx{jet}$. This is visible in the differential jet-$p_{\rm{T}}$ distribution of Fig.~\ref{fig:ptjetaj}, where the uncertainty band is particularly narrow for small transverse momenta.
Whether this additional jet veto gives rise to a sufficiently stable theoretical approximation in the sense of an experimentally applicable strategy, 
does highly dependent on the phenomenological question we ask, i.e. the phase-space region we are interested in.
Additional jet radiation, as can already be inferred from figure \ref{fig:scvarlhc}, is kinematically unsuppressed to large extent, 
especially when considering hard events with large transverse momenta. Vetoing additional radiation in a region of phase-space
where it becomes likely is crucial to the flat scale dependence of the exclusive cross section. We discuss this in more detail
in Sec.~\ref{sec:distriblhc}.

\subsection{Differential distributions at the LHC}
\label{sec:distriblhc}
We now turn to the effect of QCD corrections to the full processes 
\mbox{$pp \rig 3\,\rm{leptons}~+\dsl{E}_{\mathrm{T}} $} $+~ {\rm jet} + \mathrm{X}$.
Including the leptonic decays with the selection criteria quoted in section \ref{sec:eventsel} 
yields the cross sections given in table \ref{tab:comp} at $\mu_{\rm{R}}=\mu_{\rm{F}}=\MW$ for the inclusive and the vetoed sample, respectively.
In table \ref{tab:comp} we additionally give a precision comparison of the cross sections calculated with our 
two programs described in Sec.~\ref{sec:details}. 

\begin{table}[!b]
\begin{center}
\begin{tabular}{l  r  r   }
& Program 1 & Program 2 \\
\hline
$\sigma^{\rm{NLO}}_{\rm{incl,~decay}}({\rm{W^-Z+jet}})$    &  $ 7.4592\,[48]~{\rm{fb}}$ &$ 7.4628\,[63]~{\rm{fb}}$\\
$\sigma^{\rm{NLO}}_{\rm{incl,~decay}}({\rm{W^+Z+jet}})$    &  $11.129\,[10]~{\rm{fb}}$ & $11.1286\,[47]~{\rm{fb}}$\\[1ex]
$\sigma^{\rm{NLO}}_{\rm{excl,~decay}}({\rm{W^-Z+jet}})$    & $ 4.6721\,[62]~{\rm{fb}}$ & $4.6663\,[64]~{\rm{fb}}$ \\
$\sigma^{\rm{NLO}}_{\rm{excl,~decay}}({\rm{W^+Z+jet}})$    & $ 6.6900\,[92]~{\rm{fb}}$ & $6.6816\,[49]~{\rm{fb}}$ \\
\hline
\end{tabular}
\caption{ \label{tab:comp} Comparison of the numerical results from both of our programs to verify their excellent statistical agreement on the per-mill level for
$pp \rig 3\,\rm{leptons}~+\dsl{E}_{\mathrm{T}}+ {\rm jet} + \mathrm{X}$.}
\end{center}
\end{table}

The differences of W$^-$Z+jet compared to W$^+$Z+jet production are 
predominantly due to the different parton distribution functions of the incoming partons in the dominating subprocesses. In particular, there are no initial-state up-quarks involved in W$^-$Z$+$jet at LO, but in W$^+$Z$+$jet.

\begin{figure}[!t]
\centering
\includegraphics[bb = 200 380 450 700, scale = .8]{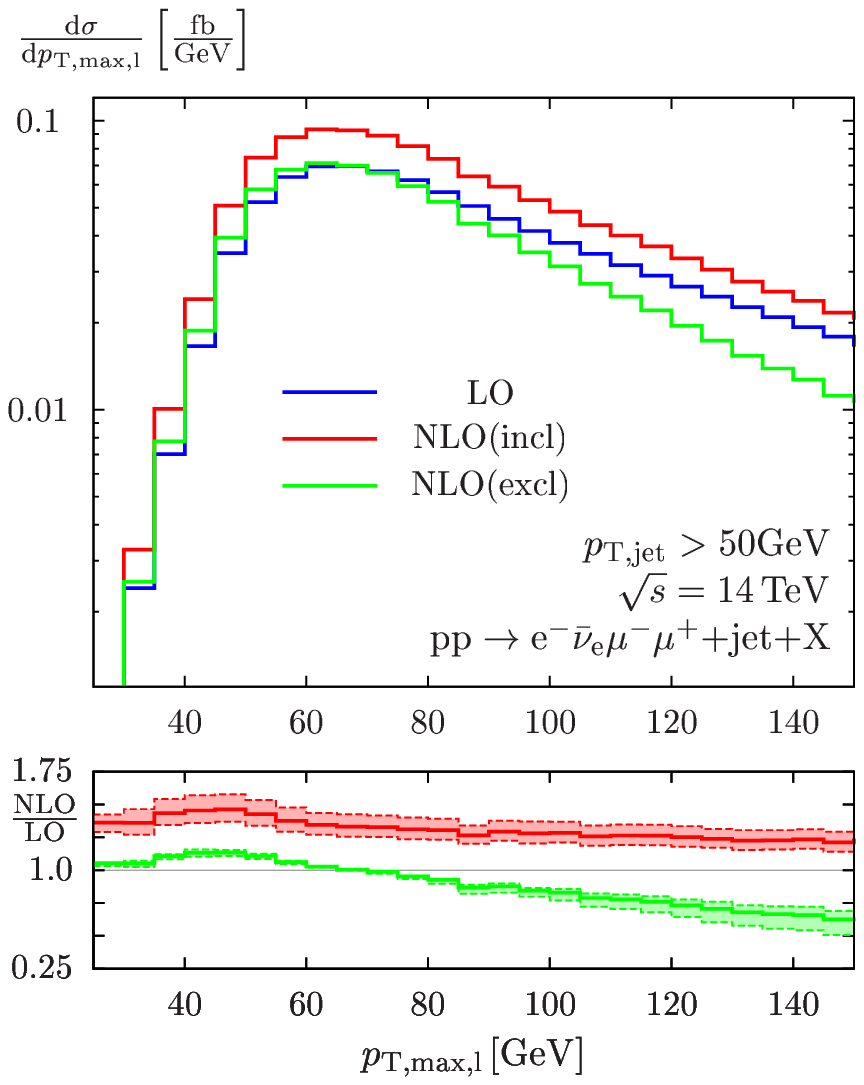}
\includegraphics[bb = 160 380 410 700, scale = .8]{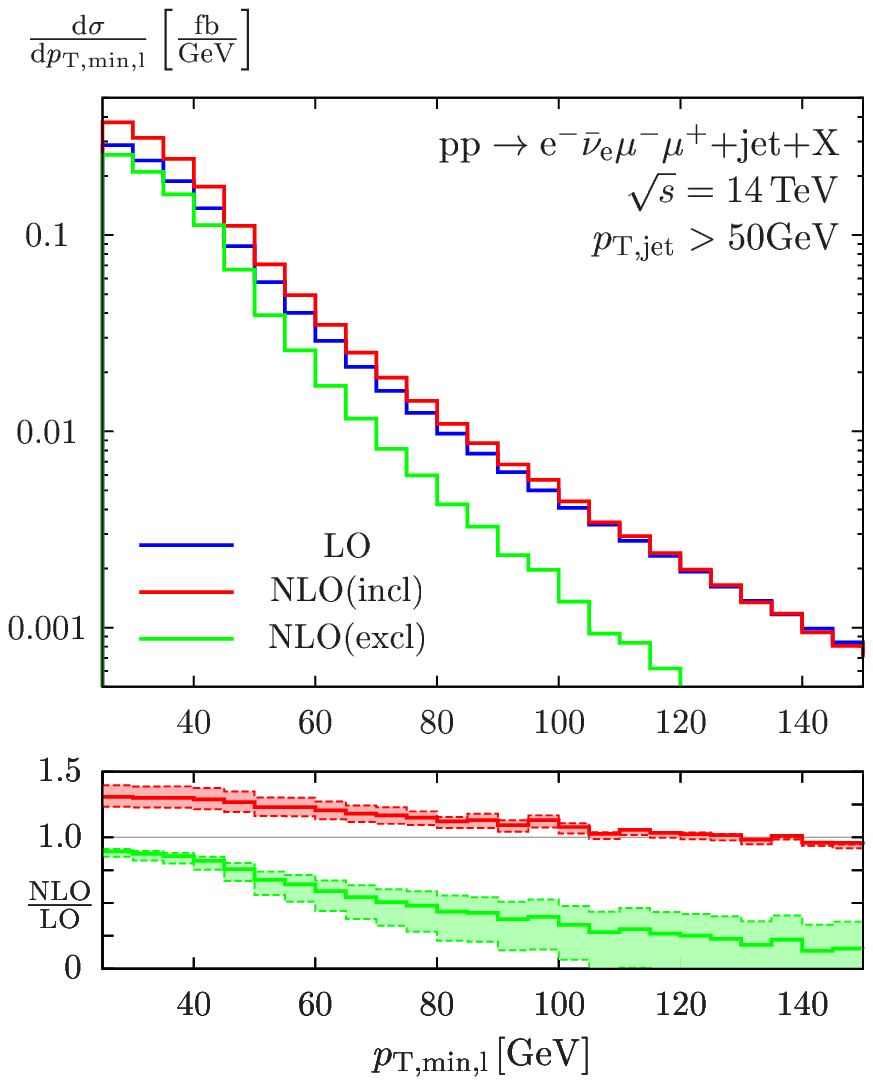}\\[-2em]
\caption{\label{fig:ptlminptlmax} LO, NLO-inclusive and NLO-exclusive differential distributions of the maximum and minimum lepton 
transverse momentum for W$^-$Z+jet production. The differential $K$-factor band corresponds to varying $\mu_{\rm{R}}=\mu_{\rm{F}}$ by a factor two around the central scale
in the NLO distribution only.}
\end{figure}

\begin{figure}
\centering
\includegraphics[bb = 200 380 450 700, scale = .8]{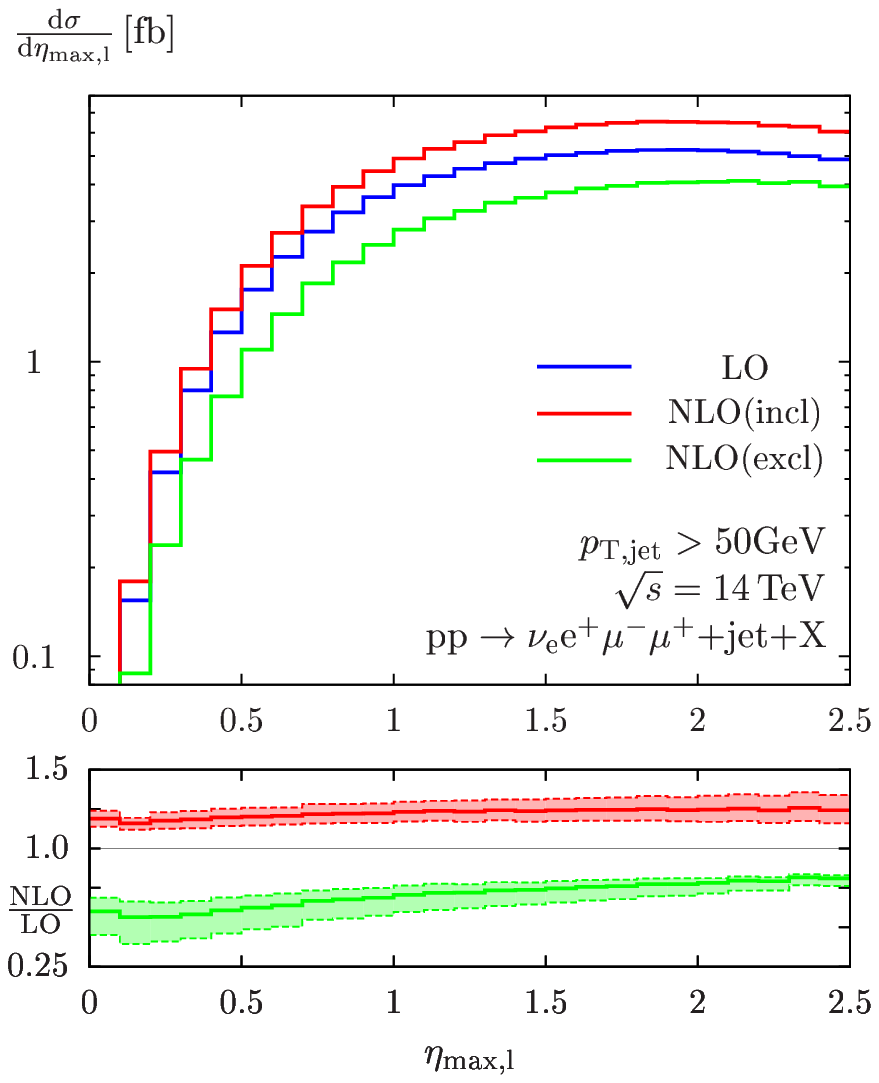}
\includegraphics[bb = 160 380 410 700, scale = .8]{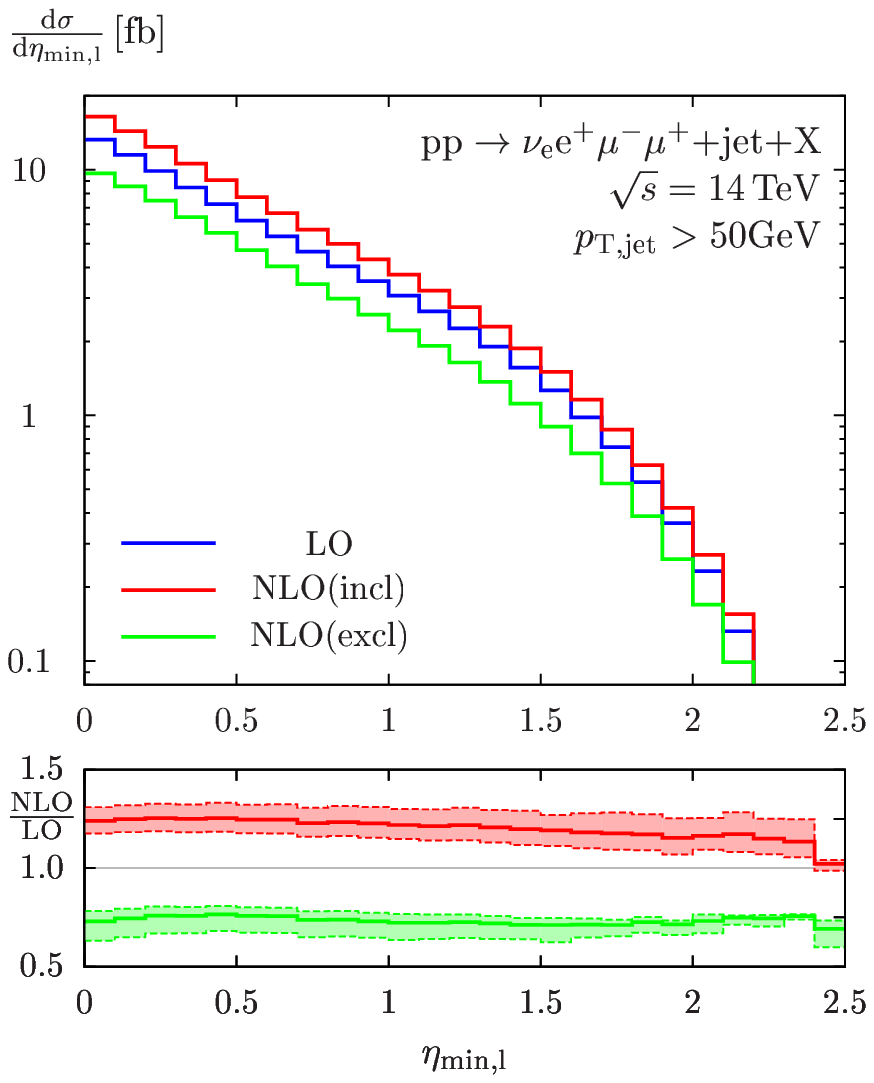}\\[-2em]
\caption{\label{fig:etatlminetalmax} LO, NLO-inclusive and NLO-exclusive differential distributions of the maximum and minimum lepton's 
pseudorapidity for W$^+$Z+jet production. The differential $K$-factor band corresponds to varying $\mu_{\rm{R}}=\mu_{\rm{F}}$ by a factor two around the central scale
in the NLO distribution only.}
\end{figure}

\begin{figure}
\centering
\includegraphics[bb = 200 380 450 700, scale = .8]{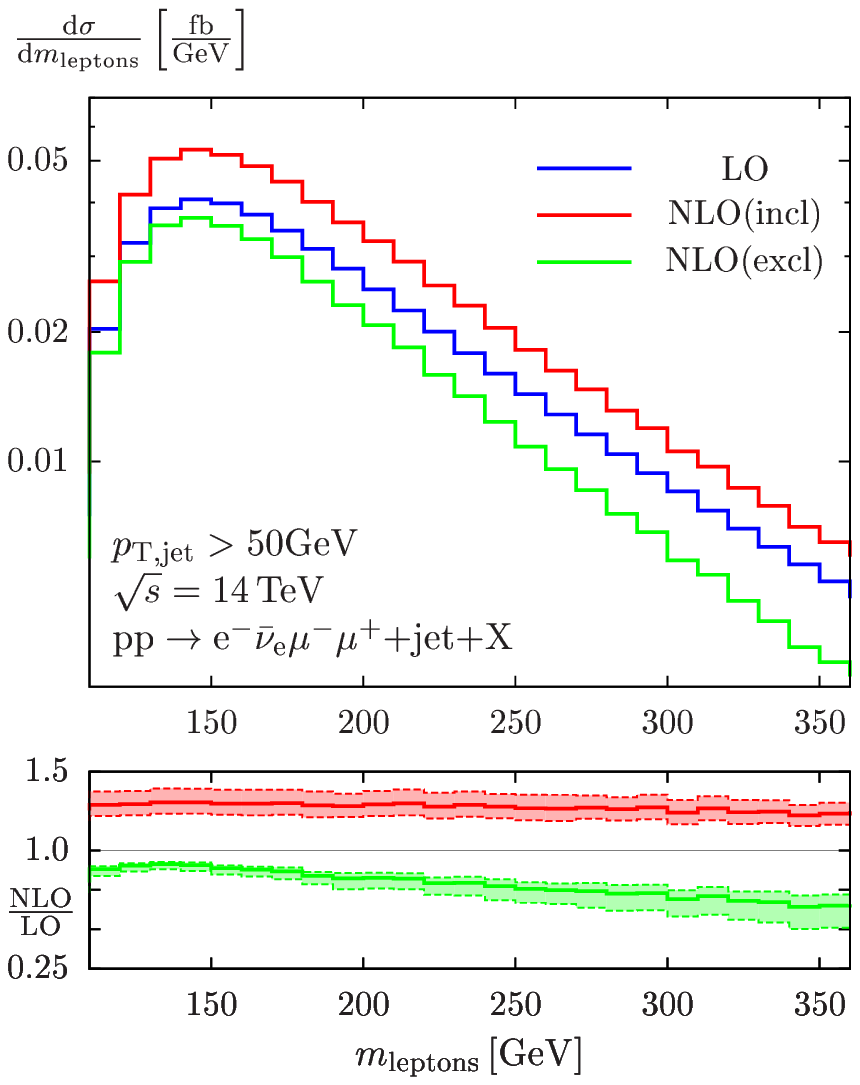}
\includegraphics[bb = 160 380 410 700, scale = .8]{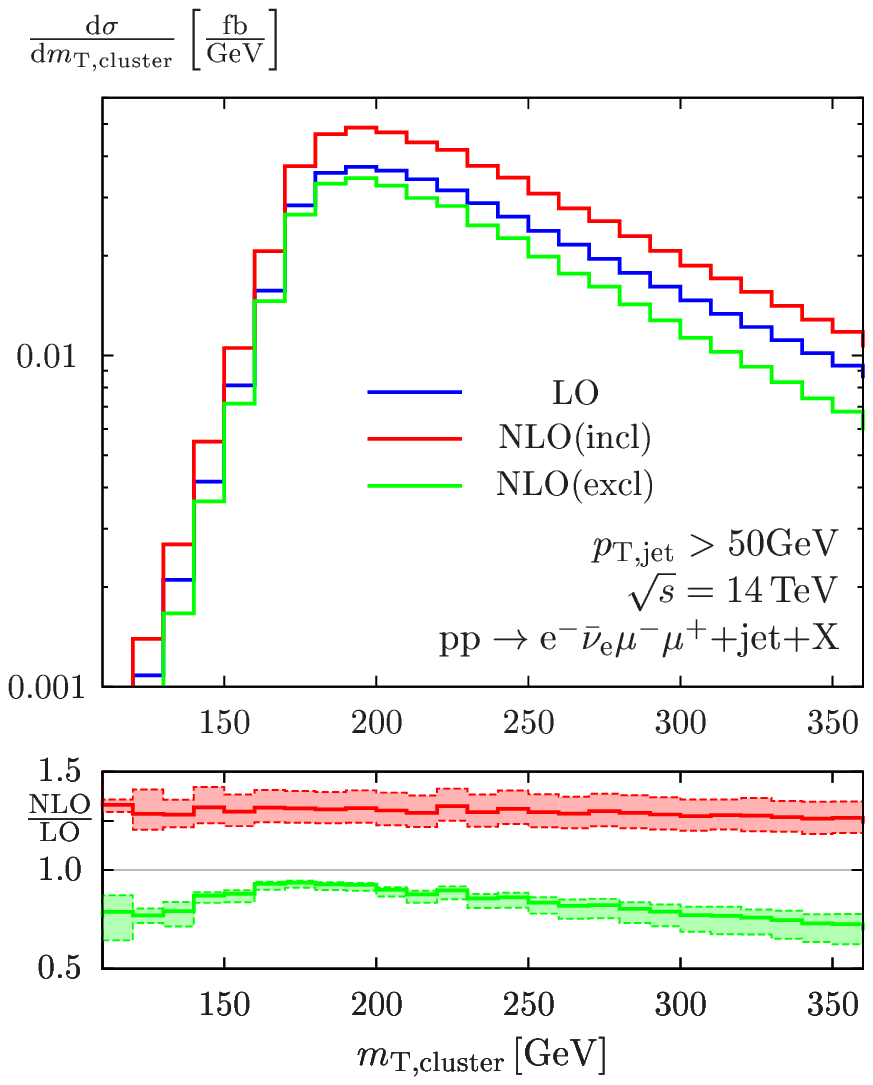}\\[-2em]
\caption{\label{fig:invlep} LO, NLO-inclusive and NLO-exclusive differential distributions of the tri-lepton invariant mass $m_{\rm{leptons}}$ and the 
transverse cluster mass for W$^-$Z+jet production. The differential $K$-factor band corresponds to varying $\mu_{\rm{R}}=\mu_{\rm{F}}$ by a factor two around the central scale
in the NLO distribution only.}
\end{figure}

From figures \ref{fig:ptjetaj}--\ref{fig:invlep} we uncover a substantial observable-specific phase-space dependence of
the QCD corrections. While additional jet radiation gives sizable contributions to the maximum-$\pTx{jet}$ distribution
at large values, the LO approximation considerably overestimates the NLO-exclusive findings. Additionally, for inclusive
events, the jets tend to be more central due to the extra hard jet emission, which occurs central at small 
rapidity differences with respect to the other reconstructed jet.

The harder inclusive jets balance against a softer inclusive lepton-$\pT$ spectrum, figure~\ref{fig:ptlminptlmax}, and additional QCD radiation.
Apart from this significant deviation of the leptonic distributions from the total $\sigma^{\rm{NLO}}/\sigma^{\rm{LO}}$-rescaling, the bulk of the leptonic
observables receive only minor differential distortions compared to rescaled-LO when including inclusive-NLO precision. 
Representatively, we show the maximum and minimum charged-lepton pseudorapidity in figure
\ref{fig:etatlminetalmax} and the tri-lepton invariant mass 
\bee
m^2_{\rm{leptons}}=(p_{e^-}+p_{\mu^+}+p_{\mu^-})^2
\eee
and the transverse WZ cluster mass in figure \ref{fig:invlep}. The transverse cluster mass
\beeq
m^2_{\rm{T,cluster}}= \left(\sqrt{m^2(e^-\mu^+\mu^-)+\vec{p}_{\mathrm{T}}^{\,2}(e^-\mu^+\mu^-)}+ |\slashed{p}_{\mathrm{T}}|\right)^2
 - \left(\vec{p}_{\mathrm{T}}(e^-\mu^+\mu^-)+\slashed{\vec{p}}_{\mathrm{T}}\right)^2\,,
\eeeq
is a convenient observable to observe production of additional charged heavy bosons \cite{Englert:2008wp,Bagger:1993zf} from
a beyond-the-SM sector via Jacobian peaks.

The exclusive distributions, even though improved perturbative stability is suggested from the decreased scale dependence of the total cross sections, which can be observed in
figure \ref{fig:scvarlhc}, exhibit large uncertainties, especially in the tails of the $\pT$ distributions.
Here additional jet radiation becomes probable as can be seen from the maximum jet-$\pT$ distribution in figure \ref{fig:ptjetaj}.
The improved NLO stability of the exclusive sample shows up as perturbative improvement almost exclusively around the threshold region. 
For the phase-space regions characterized by larger values of $\pT$, applying the additional jet veto does not yield a 
stable result anymore---at least in the chosen setup. As shown in figure \ref{fig:ptlminptlmax}, perturbative control over the exclusive production cross section is already lost at scales of about
100 GeV while the inclusive differential cross section turns out to be reasonably stable. 
Obviously the jet-veto with a fixed $p_{\mathrm{T}}$ threshold, although hinting at appealing properties by the exclusive cross sections' flat scale variations, does not easily give rise to 
a more reliable cross section prediction within the given order of perturbation theory.

\section{Summary}
\label{sec:summary}
We have computed NLO QCD cross sections and differential distributions for W$^\pm$Z production in 
association with a hadronic
jet at hadron colliders. The calculation has been extended to full leptonic final states at the LHC, 
where they are well-observable. We find the total QCD corrections to be sizeable at both the 
Tevatron and the LHC. 
At the same time they show strong phase-space dependencies in hadronic, semi-hadronic, and especially in transverse momentum distributions. 
Hence, QCD modifications should be taken into account in every phenomenological study that employs these processes.

In addition, we demonstrate that the superficial perturbative improvement for exclusive production at the LHC, which is also observed in the various other
massive di-boson+jet production cross sections \cite{Dittmaier:2007th,Dittmaier:2009un,Campbell:2007ev,Campanario:2009um,
Binoth:2009wk}, does not give rise to perturbatively stable predictions once the additional-jet veto's impact on the large $p_{\rm{T}}$ region
is taken into account. The exclusive production's reduced scale variation therefore 
expresses NLO stability for a part of our calculation which actually
is given to NLO precision. This discrimination between NLO QCD one-jet and LO di-jet contributions, which 
is inherent to fixed-order calculations, can not be carried over to experimental strategies in a straightforward way, and, hence, does not easily give rise to a 
phenomenologically applicable method.

\acknowledgments{
F.~C. acknowledges partial support by FEDER and Spanish MICINN under grant
FPA2008-02878.
C.~E. is supported by ``\mbox{KCETA} Strukturiertes Promotionkolleg" and
would like to thank the Institute for 
Theoretical Sciences at the University of Oregon for its hospitality during the time
when this work was completed. S.~K. would like to thank Stefan Dittmaier for a number of helpful discussions particularly in the initial stage of this work.
This research is partly funded by the Deutsche Forschungsgemeinschaft 
under SFB TR-9 ``Computergest\"utzte Theoretische Teilchenphysik'', and the Helmholtz alliance 
``Physics at the Terascale'', and by the EU's  Marie-Curie Research Training Network HEPTOOLS under contract MRTN-CT-2006-035505. 
}


\end{document}